# Air filtration from sarin/air mixture by porous graphene-oxide membranes: a molecular dynamics study

Marco A. E. Maria [a,b,c], Alexandre F. Fonseca [c,*]

[a] Federal University of São Carlos – Sorocaba.

[b] Facens University Center – Sorocaba.

[c] Applied Physics Department, Institute of Physics "Gleb Wataghin", State University of Campinas, Campinas, SP, 13083-970, Brazil.

**Abstract**

Sarin is a very lethal synthetic organophosphorated compound that inhibits the nervous system muscle control. Although not used as a chemical weapon anymore, it still worries the authorities regarding possible use by terrorists. Most of the studies about sarin are theoretical/computational due to its high lethality and are concentrated in its detection and degradation. Few studies are about air filtration from sarin gas. Here, the potential of graphene oxide-based membranes to filter air from sarin/air mixtures is investigated by classical molecular dynamics simulations. Membranes formed by one and two nanosheets of porous reduced graphene oxide (rGO) were considered. The passage of sarin and air molecules through these membranes from a highly concentrated region to an empty one, is evaluated as a function of temperature and sarin/air relative concentration. Sarin molecules are shown to be trapped by hydroxyl and carboxyl chemical groups in the nanosheet, while a considerable passage of air molecules ($N_2$, $O_2$ and Ar) through the membranes was verified. The results show the capacity of the rGO membranes to retain sarin from passing through, even at high temperatures, thus indicating their potential to be used as a filter for sarin gas.

**1. Introduction**

O-isopropyl methylphosphonofluoridate, better known as sarin ($C_4H_{10}FO_2P$), is one of most powerful nerve agents that worries the authorities around the World regarding its use as a chemical weapon, despite its production and stockpiling prohibition since 1997 [1–3]. The effects of sarin in human body depends on dose and time exposition, varying from vision alterations and headaches to death, basically due to the inhibition of acetylcholinesterase, which is a very important enzyme for muscles and nerves controlling, causing respiratory arrest [4,5].

Literature shows many computational works about sarin molecule, mostly based on density functional theory (DFT), and focusing on sarin adsorption on different kinds of surfaces. Dos Santos *et al*. [6], for example, reported the sarin adsorption on doped boron nitride nanotubes. Tsyshevsky *et al*. [7] and Lee *et al*. [8] studied the adsorption of sarin molecules on molybdenum dioxide and anatase titanium dioxide surfaces, respectively. Other works have been focused on the degradation of sarin molecules [9], that is predicted to start from 350 °C [9], and depends on the surface where they are adsorbed, if $TiO_2$ [10] or pristine and hydroxylated ZnO surfaces [11]. Molecular dynamics (MD) simulations and other computational techniques have been also employed in studies of sarin. Almeida *et al*. [12] and Ekström *et al*. [13] used MD simulations to study the effects of sarin on the activation of the enzyme acetylcholinesterase responsible to catalyze some neurotransmitters in the body. Mahmoudi *et al*. [14] and Lee *et al*. [15] used MD simulations to determine the diffusion coefficients of sarin encapsulated by β-cyclodextrins cavities and in polyelectrolyte membranes, respectively.

In contrast to the variety of computational works on sarin, experiments are not often found in the literature, since the manipulation of sarin gas is very dangerous and requires very controlled environments. However, few works [9,16,17] reported some experimental analyses on sarin adsorption and decomposition.

Although the relevance of the subject, there are only few studies on filtration of sarin gas [18–21], in contrast to several works on detection and degradation of this and other nerve agent gases [6-11,22–31]. On the other hand, there are many studies about the role on gas separation and detection using graphene oxide (GO) or reduced graphene oxide (rGO) [32–39]. In order to give some contribution to this topic, in this work we present a molecular dynamics based computational study about the passage of sarin/air mixture through two types of rGO membranes: (1) one formed by one porous rGO nanosheet and (2) the other formed by two parallel, non-aligned, porous rGO nanosheets. Our analyses focus on the influence of initial concentration of sarin gas, temperature, and the interaction between the gas and the membrane, on the filtration mechanisms of air from sarin gas by rGOs.

Next section presents the structural models and simulations methods. Then, the results are presented and discussed. At the end of the manuscript, the conclusion section summarizes the main results of this work.

## 2. Modeling and simulation methods

### 2.1 Modeling of the rGO nanosheet, sarin and air

The filtration system is defined by a simulation box of 34.0 Å × 32.3 Å × 200 Å of size along *x*, *y* and *z* directions, respectively, in the middle of which a membrane consisting of one or two rGO nanosheets of 32.6 Å x 30.7 Å of size, is placed orthogonally to the *z* direction of the box (see Figures 1 and 2). For the membrane with one rGO nanosheet, the simulation box is divided into two regions (I and II on each side of the membrane), with one initially containing the sarin-air mixture, and the other initially empty. For the membrane with two rGO nanosheets, the box is divided into three regions (I, II and III), where the regions I and III, on both sides of the membrane are similar to that of the previous case, and the region II is that between the two rGO nanosheets (Figure 2).

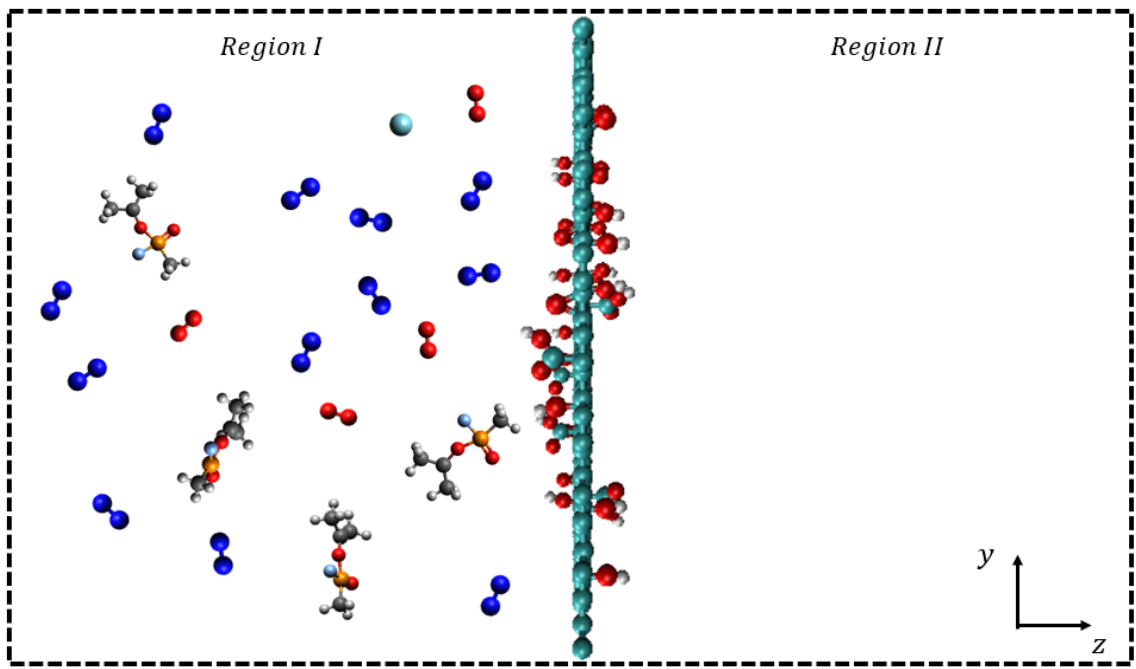

**Figure 1.** Representation in *y-z* plane of the initial configuration of the system with one nanosheet. The number of molecules drawn in the figure are not the exact ones simulated in this work and are shown here just for illustration and distinction between the two regions. Blue, red, white, orange, gray and cyan represent nitrogen, oxygen, hydrogen, phosphorous, carbon in sarin and carbon in rGO atoms, respectively.

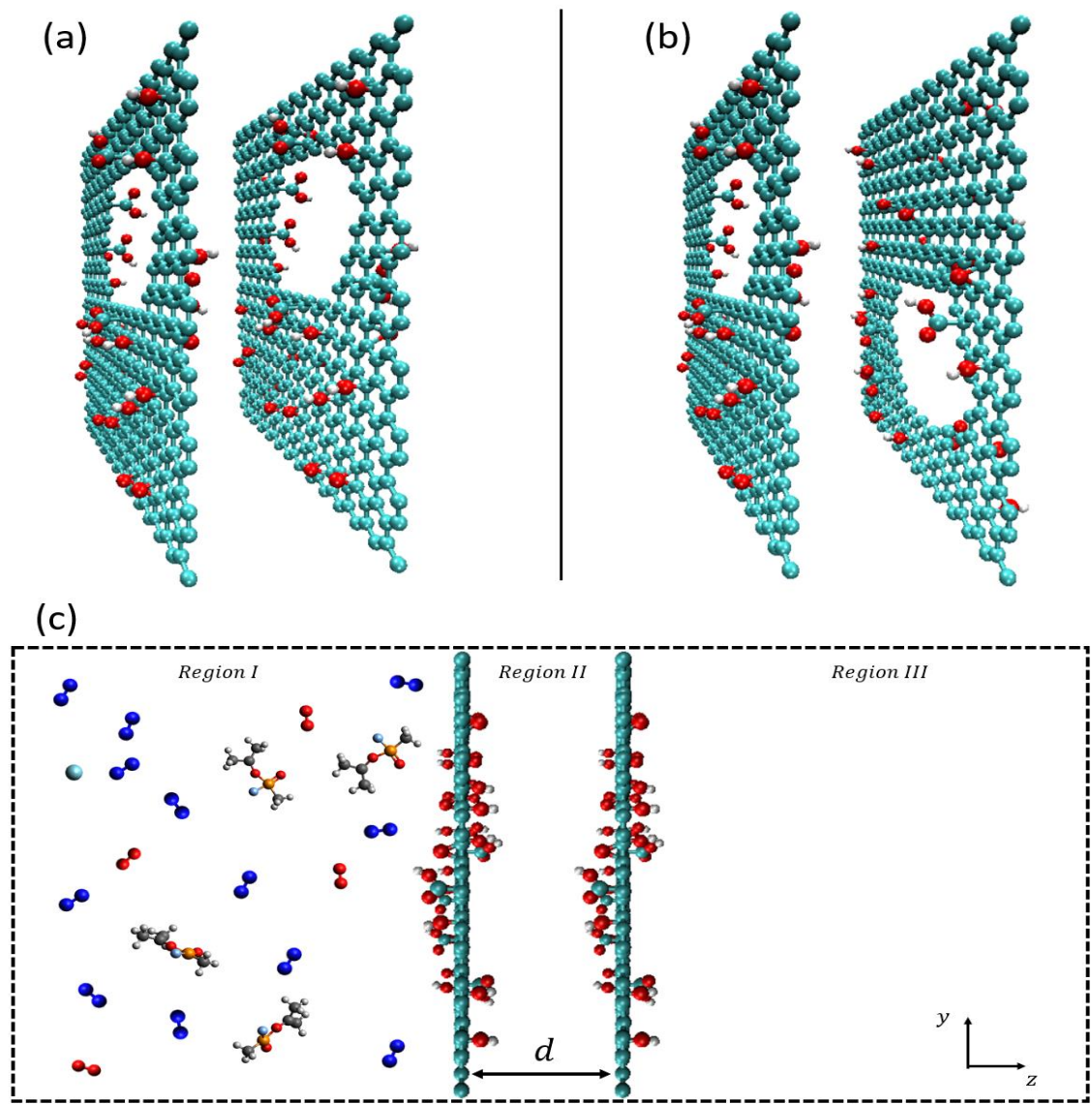

**Figure 2.** Configurations of rGO membranes with aligned (a) and non-aligned (b) nanosheet pores and the division of the simulation box in three regions (c), seen in *y-z* plane of the simulation box. The distance between nanosheets is represented by $d$. The number of molecules drawn in the figure are not the exact ones simulated in this work and are shown here just for illustration and distinction between regions. Blue, red, white, orange, gray and cyan represent nitrogen, oxygen, hydrogen, phosphorous, carbon in sarin and carbon in rGO atoms, respectively.

Figure 3 (a) shows the rGO nanosheet structure containing 9.8% of oxygenated groups: OH (hydroxyl), OCO (epoxy) and COOH (carboxyl). On the rGO plane, hydroxyl and epoxy groups are randomly distributed while carboxyl groups are placed at the edges of a pore of about 193.4 Å² of area. The concentrations of hydroxyl, epoxy and carboxyl groups with respect to carbon atoms of the nanosheets are 3.7%, 4.0% and 1.5% as considered in previous studies [40–42]. The sarin molecule model considered here is based on references [43,44], that used the united-atom approach for the $CH_3$ groups, while all atoms were explicitly considered for the rGO structure models. Figure 3 (b) shows the sarin molecule with atomic partial charges. The pore size, presented in Figure 3 (a), was chosen to allow the passage of a sarin molecule as a whole.

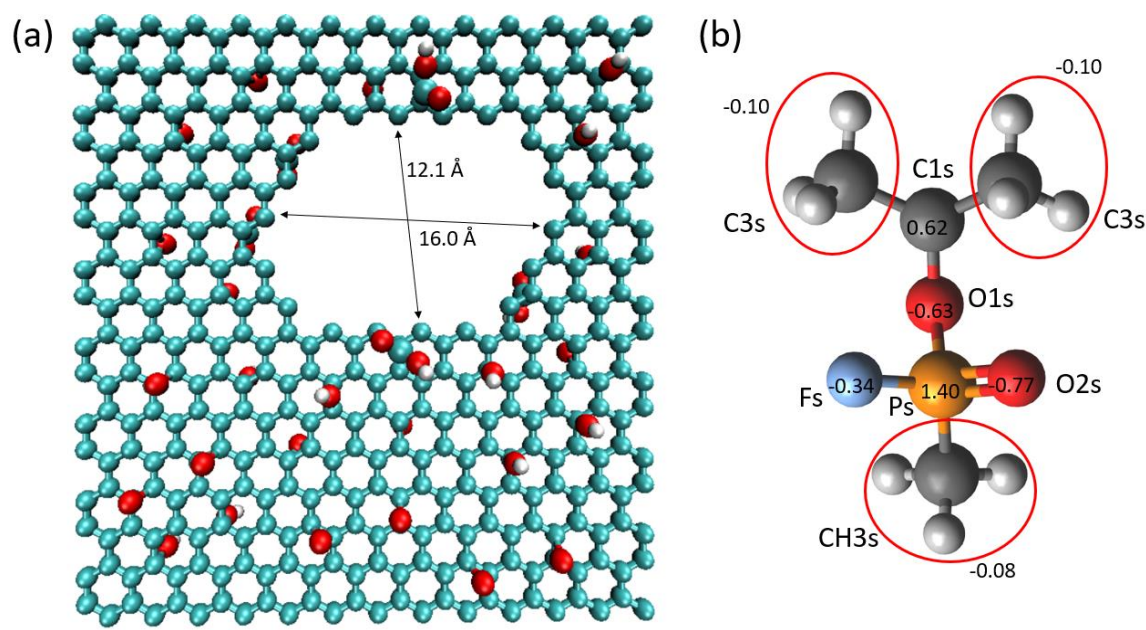

**Figure 3**. Porous rGO nanosheet (a) and sarin molecule (b) models used in this work. In (b), the values of the electrical charge of each atom is shown, in units of $e$. Blue, red, white, orange, gray and cyan represent nitrogen, oxygen, hydrogen, phosphorous, carbon in sarin and carbon in rGO atoms, respectively. $CH_3$ groups are circulated in red and their names "C3s" and "CH3s" distinguish their relative position on the sarin molecule.

For air modeling, it is adopted the composition of 78% of nitrogen molecules ($N_2$), 21% of oxygen molecules ($O_2$) and 1% of argon atoms (Ar) [45].

To describe the intramolecular interactions, the following potential energy function is used:

$$U_{intra} = k_B(r - r_0)^2 + k_\theta(\theta - \theta_0)^2 + U_{dihedral}. \qquad (1)$$

The first two terms in this equation refer to bond distance and bond angle energies, respectively, and the last term describes the torsions. $k_B$ and $k_\theta$ are constants related to the magnitude of the corresponding interactions; $r$ and $\theta$ represent, respectively, instantaneous bond distance and bond angle while the index "0" indicates the equilibrium distance ($r_0$) and angle ($\theta_0$).

The potential associated to dihedral torsions of the structure is system dependent and is given by:

$$U_{dihedral} = \begin{cases} k_d[1 + \cos(n\phi - d)], \text{ for the rGO;} \\ \sum k_i[1 + \cos(n_i\phi - d_i)], \text{ for dihedrals A, B and C in sarin;} \\ \sum A_m \cos^{m-1}(\phi), \text{ for dihedral D in sarin,} \end{cases} \qquad (2)$$

where A, B, C and D represent the dihedral torsional angles associated to the following sequences of atoms in sarin: $C1s - O1s - Ps - O2s$, $C3s - C1s - O1s - Ps$, $C1s - O1s - Ps - CH3s$ and $C1s - O1s - Ps - Fs$, respectively. In the above equation, $\phi$ is the torsional angle, while $d$ and $d_i$ are phase factors; $k_d$, $k_i$ and $A_n$ are related to the magnitude of the torsion barrier energy; $n$ and $n_i$ are the multiplicity (number of potential minima generated in a 360° rotation).

Completing the description of the potential energy of the whole structure, the intermolecular interactions are here represented by the following van der Waals and Coulombic general terms:

$$U_{inter} = 4\varepsilon\left[\left(\frac{\sigma}{r}\right)^{12} - \left(\frac{\sigma}{r}\right)^6\right] + \frac{q_i q_j}{\epsilon r}. \qquad (3)$$

The parameters for rGO, air and sarin models were obtained from references [42,46], [42,47] and [18,43,44], respectively.

*2.2 Simulation details of the membrane system with one rGO nanosheet*

The rGO nanosheet in the one membrane system is fixed at the middle of the simulation box corresponding to the position $z = 103$ Å. As previously mentioned, it divides the simulation box in two regions, here named I and II as shown in Figure 1. The sarin/air mixture is initially placed on region I, while region II is empty. MD simulations of sarin/air mixtures were carried out for 300 K, 500 K, 700 K and 900 K, during 10 ns, using a timestep of 1.0 fs. For each temperature, different sarin concentrations were considered as follows: the number of sarin atoms is always kept constant and equal to 20 molecules, but the number of air constituent molecules, $N_2$, $O_2$ and Ar, varied so as to obtain the sarin concentrations of 9%, 17% and 28%. The number of sarin and air molecules were defined to be significant to counting molecules that pass through the membranes. At the same time, the number of gas molecules together with the choice of the box size, imply a situation in which the pressure difference between the regions on both sides of the membranes is very high, as compared to the atmospheric pressure (about 25 atms for 28% of concentration of sarin), in order to verify the rGO membrane potential for air filtration from sarin/air mixture in a situation of high pressure gradient. All the simulations were carried out using classical molecular dynamics implemented through the LAMMPS package [48], on a constant number of molecules, volume and temperature (NVT) ensemble, with Nosé-Hoover algorithm for temperature control and periodic boundary conditions along $x$ and $y$ directions. Along the $z$ direction, at the edges of the simulation box, a repulsive potential is applied to the system to ensure that the only way a molecule or atom goes from region I to II or back is through the pore in the nanosheet.

*2.3 Simulation details of the membrane system with two rGO nanosheets*

Two configurations of the membrane formed by two rGO nanosheets are considered in this part of the study: i) one with aligned pores; and ii) the other with non-aligned pores. Figures 2 (a) and (b) show these configurations, respectively. For each configuration, three values of the distance, $d$, between the nanosheets were considered, ($d$ = 7 Å, 10 Å and 13 Å), thus creating three regions, as already mentioned before (in subsection 2.1) and shown in Figure 2 (c): Region I - where sarin/air mixture is initially placed; Region II - an intermediate region between the two nanosheets; and Region III - an initially empty region. For this system, we performed simulations at 300 K, 600 K and 900 K, also using timestep of 1.0 fs, and for a total amount of simulation time of 15 ns. In all these simulations, the sarin concentration was constant and equal to 17% (considering 20 sarin molecules). For the simulations with one and two nanosheets, the properties were evaluated using the data from the last 1.0 ns.

## 3. Results and discussion

*3.1 Analysis for one nanosheet membrane*

Table 1 shows the number of sarin molecules, $N$, that passed through the pore of the system with just one rGO nanosheet for different temperatures and sarin concentrations, after 10 ns of simulation. It is possible to see in this table that both sarin concentration and temperature influences the values of $N$. For low temperature, however, it is remarkable to see that after 10 ns, not a single sarin molecule has reached the region II, despite the high pressure difference. It is important to highlight that the pressure at the initial sarin/air mixture region I varied from 79 to 25 atms, respectively to the sarin concentration variation from 9% to 28%. As the temperature increases, the passage of sarin molecules increases as well, where it is also possible to see the dependence of $N$ with the initial sarin concentration. The dependence of $N$ on temperature is coherent with the fact that higher values of kinetic energy (directly affected by temperature), increases the chances of a sarin molecule to cross from region I to region II. As we have fixed the number of sarin molecules in all simulations, and sarin concentration is changed by changing the number of air molecules, small sarin concentration means the presence of more air molecules within the system and, therefore, more chances of air molecules to occupy the space close to the rGO pore and, consequently, more chances of sarin molecules to collide with them, thus helping to avoid their passage through the pore.

**Table 1.** Number of sarin molecules, $N$, that move from region I to region II through the pore after 10 ns of MD simulation at four temperatures (300 K, 500 K, 700 K and 900 K). The number of sarin molecules initially confined in region I was constant and equal to 20.

| Sarin concentration [%] | $N$ (300 K) | $N$ (500 K) | $N$ (700 K) | $N$ (900 K) |
|---|---|---|---|---|
| 9 | 0 | 0 | 4 | 7 |
| 17 | 0 | 4 | 8 | 6 |
| 28 | 0 | 6 | 4 | 10 |

Another point that was observed from the simulations is that for all conditions, even for high temperature simulations, sarin molecules always end up surrounding the rGO nanosheet. This can be observed in Figure 4, which presents the radial distribution function, g(r), between the carbon atom (C) from rGO nanosheet and central oxygen atom (O1s) from sarin, for each value of sarin concentration studied here (Figure 4 (a)) and each value of the simulated temperature (Figure 4 (b) for 17% of sarin concentration). In both graphics, the presence of a peak at about 7 Å reveals a considerable attraction between sarin and the nanosheet, that remains the same despite the different simulated temperature or sarin concentration in the system.

We analyzed in more detail the origin of this attraction and we have found, based on the presence of the peaks in radial distribution function curves (whose additional graphs are shown in sections A, B and C of the Supplementary Materials), that it comes mainly from three interactions: one between the hydrogen atom of carboxyl group ($H_{carb}$) and the O2s oxygen from sarin (see Figure 2(b)); the other between hydrogen from hydroxyl ($H_{hyd}$) and the same O2s oxygen atom; and the interaction between the hydrogen atom from carboxyl and fluorine atom from sarin (Fs). These latter two interactions can explain why, in Figure 4 (b), an increasing temperature up to 700 K causes an increasing in g(r) values around the peak, and for 900 K, the value decreases to below that of 300 K. Since O2s and Fs atoms are oppositely separate by the phosphorous (Ps), as the temperature increases, the displacement of sarin molecule, caused by thermal agitation, can favor the approximation between Fs and hydroxyl group. At 900 K, the thermal agitation is high enough to give energy to some sarin molecules to move away from the nanosheet, thus decreasing the peak value of g(r). Figures 4 (c) and (d) show g(r) for $H_{hyd}$ – O2s and $H_{carb}$ – O2s at 300 K and 900 K, all for sarin concentration equal to 17%. Inset in Figure 4 (d) highlights the peaks about 2.0 Å of the g(r) function corresponding to the $H_{carb}$ – Fs pair of atoms, also at 300 K and 900 K. Again, these results evidence that the attraction between rGO and sarin keeps strong even at high temperatures (g(r) curves for 500 K and 700 K were not shown in Figures 4 (c) and (d) for the sake of clarity, but they are shown in Supplementary Material, in section D).

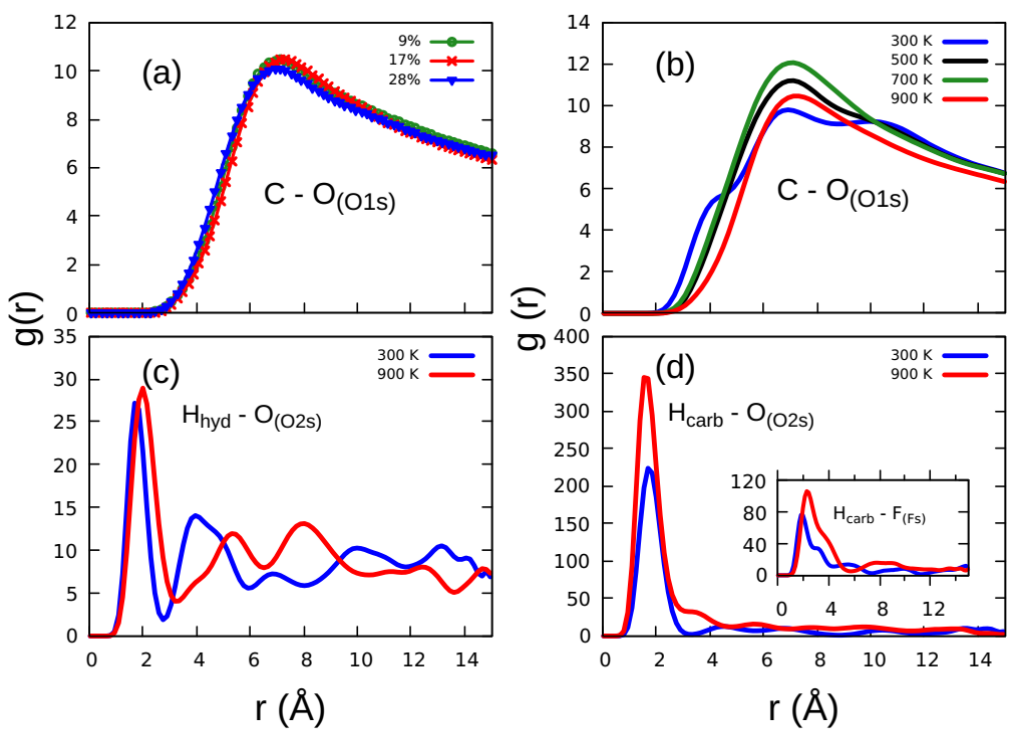

**Figure 4.** Radial distribution functions, g(r), between carbon atom from nanosheet (C) and sarin central oxygen (O1s) as function of concentration (a) and temperature (b), and between hydrogens, from hydroxyl (c) and carboxyl (d) groups, and the oxygen (O2s) of sarin. The inset depicts the g(r) resulting from the interaction of hydrogen from carboxyl and sarin fluorine atom. All curves shown in graphs in (b), (c) and (d) corresponds to 17% of sarin concentration.

In order to further understand the attraction between sarin and rGO nanosheet, we have evaluated the potential energy of one sarin molecule as a function of the distance to the nanosheet, considering three possible molecule orientations relative to the plane of the nanosheet: perpendicular, 45° tilted, and parallel to the nanosheet plane. The molecule is quasi-statically moved by small increments of 0.05 Å from –10 to 10 Å keeping the atoms of the rGO fixed. The energy curves for the three orientations are presented in Figure 5 (which also shows the sarin orientations in relation to rGO as well), where the $z$ coordinate of the nanosheet was set to $z = 0$ Å. For all sarin orientations, one can notice the presence of at least one potential well with minimum value around the distance of 2 Å from the nanosheet carbon plane. For the perpendicular orientation of sarin, the potential presents an additional minimum around 3.7 Å. The asymmetry along *z* in the format of the potential curves come from the fact that functional groups in rGO are not symmetrically distributed over both of its sides, causing a natural asymmetry in the distances between each atom of the sarin molecule and the functional groups in the nanosheet. We compare the values of the potential energy associated to each depth of these potential energy minima to thermal energy $k_B T$ ($k_B$ is the Boltzmann constant) at room temperature ($k_B T \sim 0.59$ kcal/mol), and both values for each minimum in the three graphics. These results show that sarin molecules need high kinetic energy to escape the rGO attraction. They also explain the absence of dependence of the peak position of the radial distribution functions, g(r), on sarin concentration or temperature (up to 900 K) shown in Figure 4.

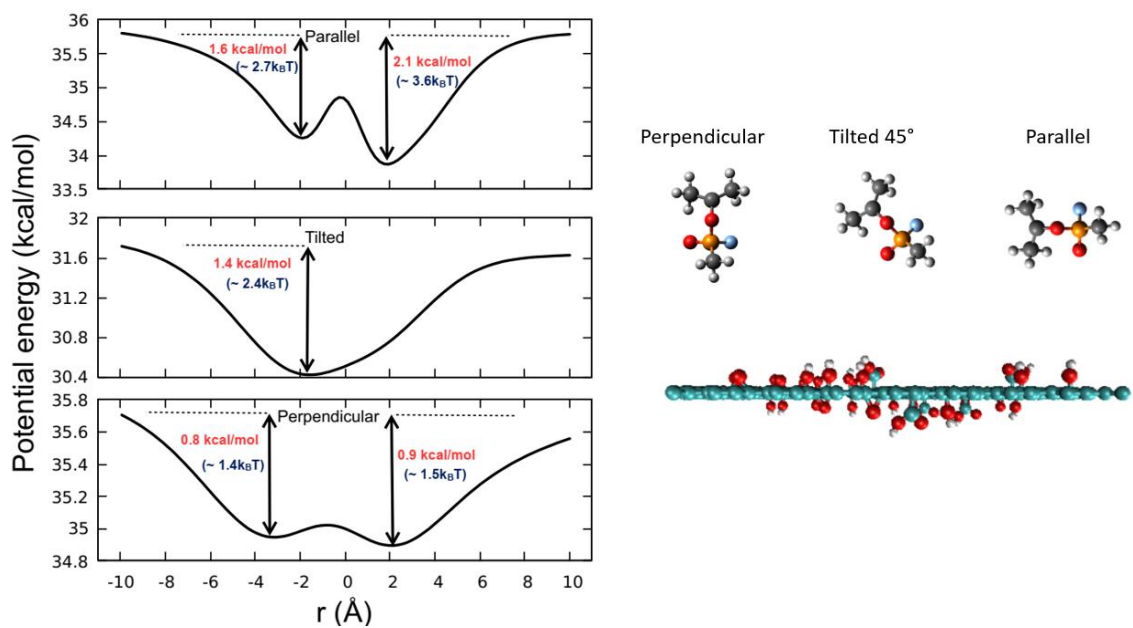

**Figure 5.** Potential energy curves for three different orientations (shown aside) of sarin relatively to carbon plane of the nanosheet. In details are shown the estimated values of each potential well in kcal/mol (in red) and in terms of the thermal energy, $k_B T$, at room temperature (in blue).

Following these observations, it is also important to compare the distribution of sarin and air molecules around the rGO nanosheets, i.e, along the *z* coordinate of the simulation box. In order to do so, the number of sarin and air molecules per volume, *N/V,* along the *z* direction, in the equilibrium, was calculated as follows. The simulation box was divided into slabs and the number of each type of molecules lying in each slab was counted and averaged along the last 1 ns of simulation. Figures 6 (a) and (b), then, show the concentration of sarin and air molecules along the *z* coordinate, respectively. Figure 6 (a) shows that closer to the nanosheet, larger the number of sarin molecules, for different temperatures and 17% of sarin concentration. Figures 6 (a) and 6 (b) show that the largest values of *N/V* of both air and sarin molecules around the membrane (around $z \sim 0$ Å) happen for $T = 300\ K$. This is justified by the fact that at this temperature, the kinetic energy of sarin and air molecules is not enough to push them away from the nanosheet. But at different temperatures, the concentration of sarin molecules around the membrane monotonically decreases with temperature while the air concentration does not. Although most of the air molecules also surround the nanosheet, it is clear that sarin gets much more concentrated around the rGO than air molecules, which get spread out amongst the two regions I and II. This confirms the potential energy analysis between sarin and rGO shown in Figure 5 and reveals that the air-rGO attraction is less effective than that between rGO and sarin, even for high temperatures. Similar results were reported for the system containing a mixture of sarin and air crossing a pore on a graphene nanosheet [18].

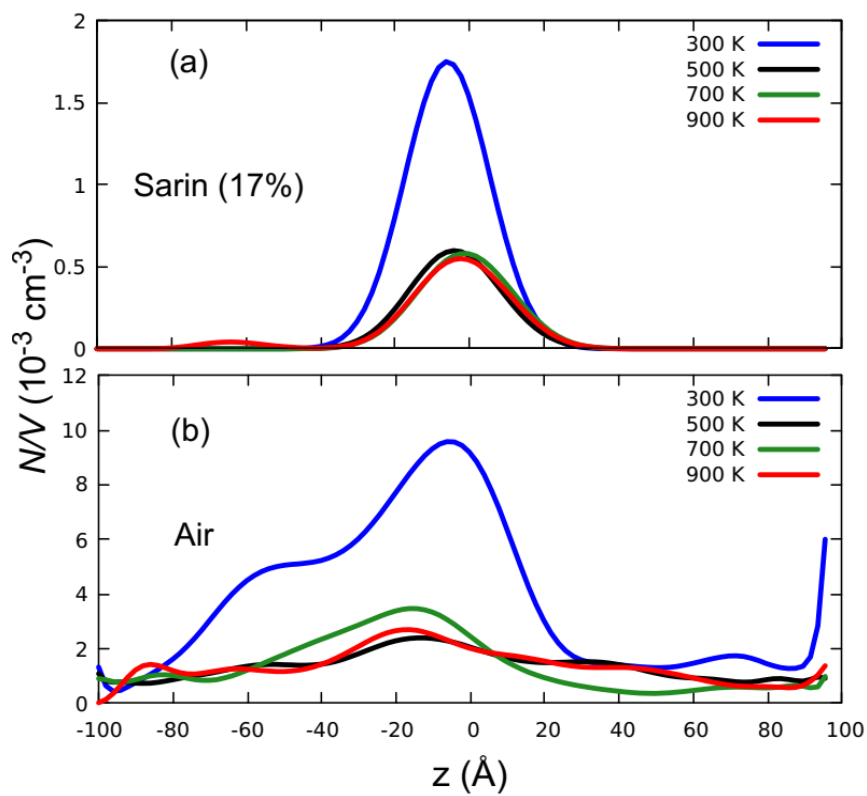

**Figure 6.** Density number of molecules along $z$ direction, of sarin (a) and air (b) for different temperatures.

*3.2 Analysis for two nanosheets membranes*

The analyses performed for the systems with two rGO membranes are similar to that performed for the system with just one nanosheet. The number of sarin and air molecules that moved from the region I to regions II and III, for different rGO distances ($d$ = 7, 10 and 13 Å), for different temperatures ($T$ = 300, 600 and 900 K) and for aligned and non-aligned orientations of pores were investigated for the particular case of 17% of sarin concentration. The results are summarized in Figure 7, which shows the number of sarin molecules present on regions II (Figure 7 (a)) and III (Figure 7 (b)) at the end of 15 ns MD simulations for each combination of $d$ and $T$ values, and one particular result for a test of a long 60 ns MD simulation at $d$ = 13 Å and $T$ = 600 K. Figure 7 (a) shows that the number of molecules in region II (between the nanosheets) is, in general, higher for the configuration of aligned pores than for non-aligned ones. It is also observed that the number of sarin molecules in each region is affected by both the distance between nanosheets and the temperature of the system. The effect of temperature is obvious, since larger the average kinetic energy of sarin molecules, higher the chance to escape from the potential energy attraction of rGOs. Increasing $d$ increases the volume of the region II, which provides more room to accommodate more sarin molecules. Therefore, larger the distance, $d$, larger the number of sarin molecules in region II and, then, higher the chance for a sarin molecule to reach the region III. According to Figure 7 (a), there is no much difference in the number of sarin molecules in region II between systems with membranes formed by aligned and non-aligned rGO pores, if the temperature is smaller than 900 K. Except for the membrane configuration of aligned rGOs at $d = 7$ Å, for which a single sarin molecule reached the region III after 15 ns of simulation, Figure 7 (b) shows that none sarin molecule reached region III for both membrane configurations at temperature values up to 600 K after 15 ns of MD simulations. At 900 K, however, Figure 7 (b) shows that the non-aligned pores membrane configuration was able to completely block the passage of sarin for all conditions of temperature and rGOs distance, except for $d = 13$ Å. At this last value of rGOs distance, some sarin molecules had got enough energy to escape the membrane attraction and reach region III.

In order to verify if the filtration of air from sarin molecules by the two rGOs membranes is not a time-dependent result, we performed MD simulations for up to 60 ns (four times longer than previous ones) for both aligned and non-aligned membrane configurations for the particular case of $d = 13$ Å and $T = 600$ K. Figures 7 (a) and (b) show the results for the number of sarin molecules in regions II and III, respectively, after 60 ns of MD simulations for $d = 13$ Å and $T = 600$ K, for both aligned and non-aligned pores configurations. It is observed that six sarin molecules moved to region II for the aligned configuration (Figure 7(a)) and none to region III (Figure 7 (b)). This suggests that the concentration of sarin molecules around and between the two rGO nanosheets is a steady state for temperatures up to 600 K.

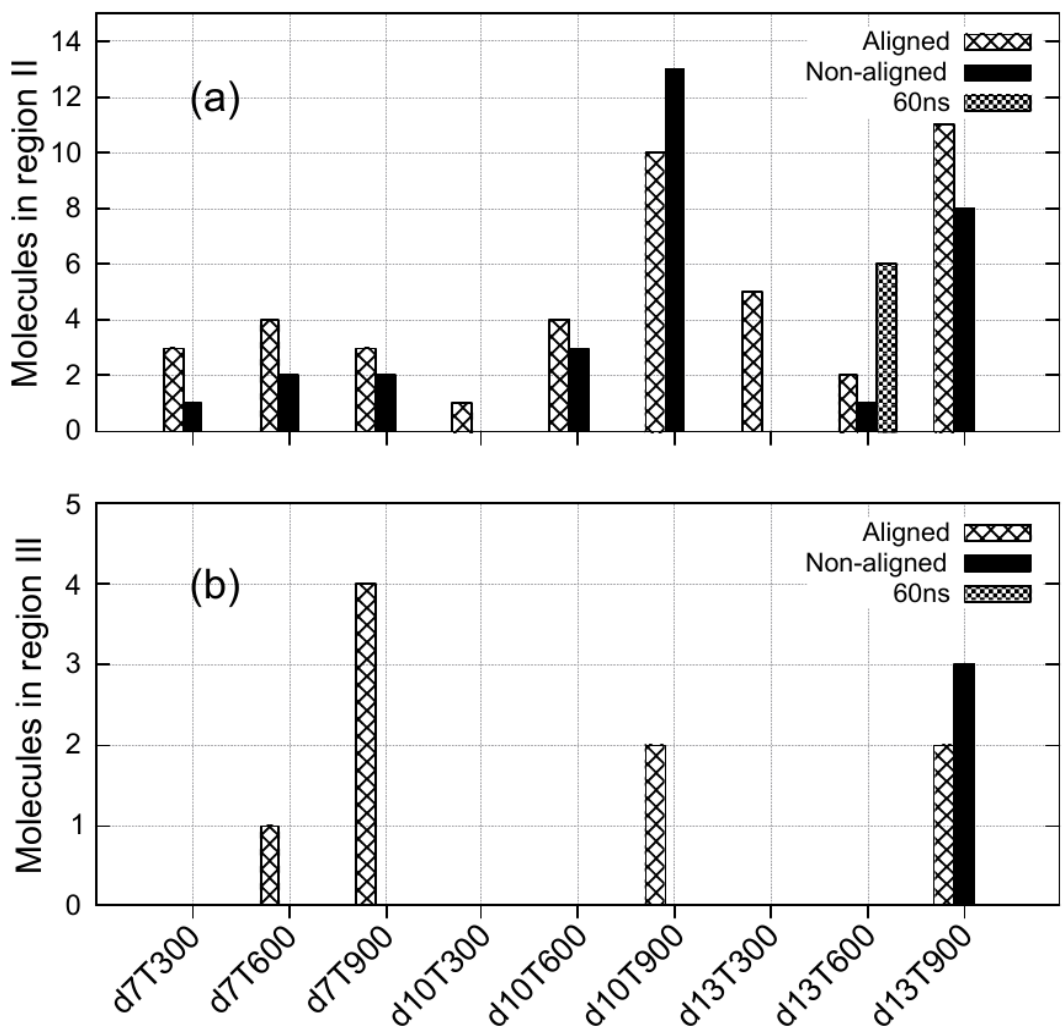

**Figure 7**. Number of sarin molecules in region II (a) and region III (b) for aligned and non-aligned pores membrane configurations after 15 ns of MD simulations. For the particular case corresponding to $d = 13$ Å and $T = 600$ K, for which no sarin molecules were found in region III after 15 ns, additional 45 ns of MD simulations were performed for both aligned and non-aligned configurations. The number of sarin molecules after 60 ns are shown in panels (a) and (b).

Figure 8 shows the distribution of air molecules along the $z$ direction for membranes of aligned and non-aligned rGO pores after 15 ns of MD simulations. The position of the first nanosheet in all graphics is set to $z = 0$ Å. Regardless of the pore alignment, Figures 8 (a) and (c) show, for fixed $d = 7$ Å (region II extends, in these two graphics, from $z = 0$ Å to $z = 7$ Å) and temperatures greater than 300 K, a shift of the distribution curves to the left, i.e., towards the $z < 0$ region (region I) and, for the aligned case, a small increasing of absolute values of the number of air molecules for $z > 30$ Å (region

III). These results show that some air constituent molecules that cross the two rGO nanosheets have enough energy to reach region III, despite most of the air molecules tend to wander near to nanosheet. The shift to the left of the peaks of the air distribution curves for temperatures of 600 K and 900 K in comparison to that of 300 K, can be explained by the following: the air molecules have their kinetic energy increased with increasing temperature. It gives more mobility to the air molecules to move around and along the simulation box. As the membrane imposes a barrier against crossing to region III, the movement within region I at distances far from the first rGO also increases, thus causing the decrease of the number of molecules that surround the nanosheet (decrease of the magnitude of the peaks for $T > 300$ K). At the same time, for the small number of air molecules that reach the region II, increasing the temperature causes the increase of the chance of these molecules to pass through the second rGO nanosheet thus reaching the region III. That is why the number of air molecules for $T > 300$ K that are seen in region III, at $z > 30$ Å, approximately doubles when compared to that of $T = 300$ K, at least for the aligned configuration. For the non-aligned configuration, the increase of temperature is not enough to increase the passage of air molecules to region III, so there are more air molecules spread out region I for the non-aligned configuration than for the aligned one. The distance between the nanosheets does not influence much the number of molecules around the first nanosheet for both pore orientations, as can be seen by the similarity of the position of all peaks in Figures 8 (b) and (d), for $T = 300$ K and different values of $d$ (7 Å, 10 Å and 13 Å). However, there is a small decrease in the magnitude of this peak for $d = 13$ Å for both aligned and non-aligned configurations. For $d = 13$ Å in the aligned configuration, the decrease of the value of the peak is compensated by a slight increase of the number of the air molecules that passed to the region III (red curve in Figure 8(b)). For $d = 13$ Å in the non-aligned configuration, there is an increase in the number of air molecules in region I (red curve in Figure 8(d)). Although non-aligned configurations block much more air molecules than the aligned configurations, the first did not prevent them to reach the region III. The remained air molecules in region III for the non-aligned configurations are such that the pressure in region III is about 16,5 atms. That is an important result when the number of sarin molecules are analyzed in the same way.

Figure 9 shows the distribution of sarin molecules along the $z$ direction after 15 ns of simulations. Small shifts to the left and to the right of the peak of this distribution can be seen for different values of $d$ and $T$. Although the similarity between sarin and air regarding the accumulation of molecules around the membranes, there is an important difference that makes the rGO membranes potentially appropriated to filter air from sarin gas. For all studied values of $d$ and $T$, the number of sarin molecules in region III far from the membranes are null. The small shifts to the right of the peaks of the curves in Figure 9 for aligned configurations when compared to that of non-aligned ones, show that although the difference is relatively small, non-aligned configurations block more sarin molecules than aligned ones. Differences in the shape and size of the distributions shown in Figures 8 and 9 might come from the fact that sarin size and attraction to rGOs decrease its mobility as compared to that of the air.

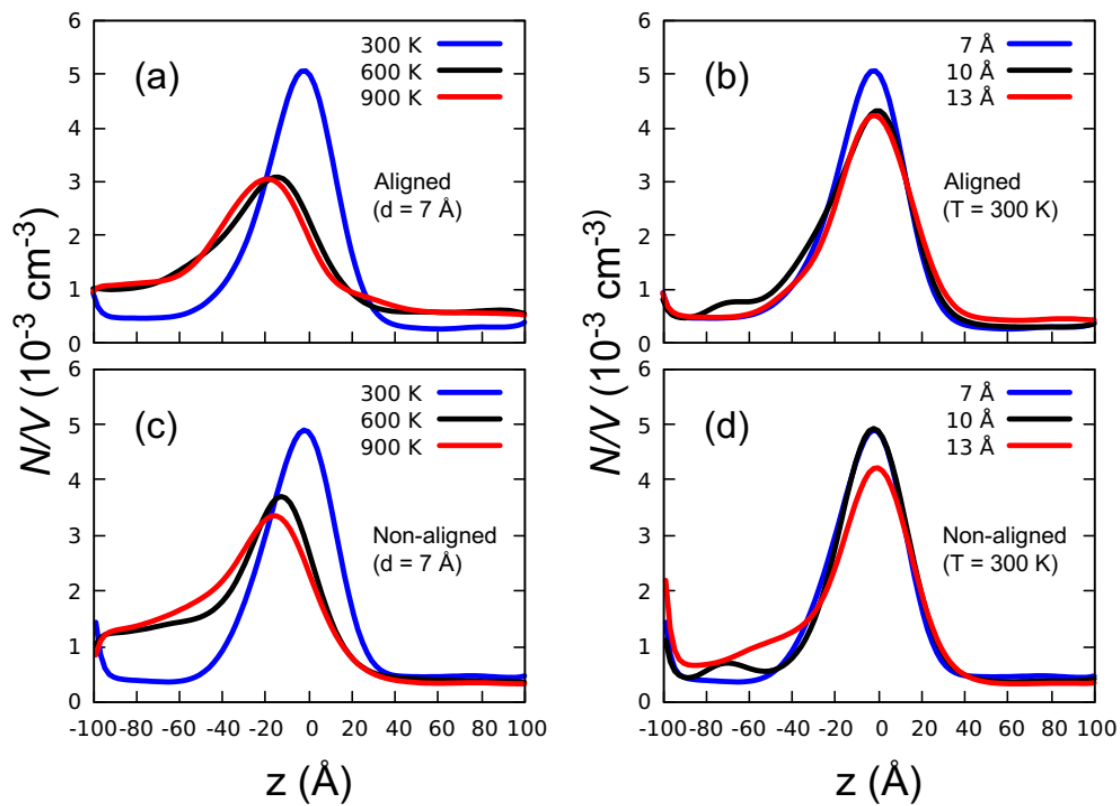

**Figure 8.** Distribution of air molecules (*N/V*) along *z* direction for different temperatures and distances between the nanosheets for aligned, (a) and (b), and non-aligned, (c) and (d), membrane configurations.

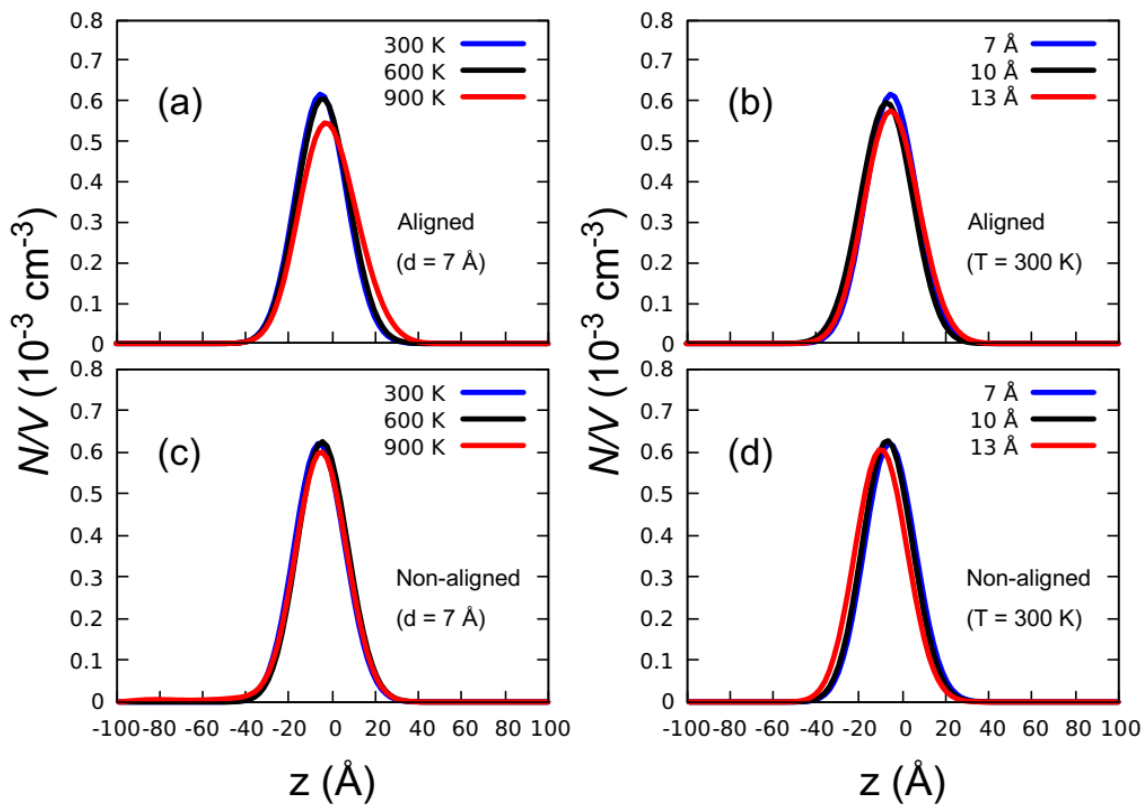

**Figure 9.** Distribution of sarin molecules along $z$ direction for different temperatures (panels (a) and (c)) and distances (panels (b) and (d)) between the nanosheets for aligned, (a) and (b), and non-aligned, (c) and (d), membrane configurations.

## 4. Conclusions

In this work, classical molecular dynamics simulations were performed to evaluate the ability of some models of graphene oxide-based membranes to filter air from a sarin/air mixture. Membranes with one or two porous rGO nanosheets were considered and the results are summarized as follows. Membranes with one porous rGO nanosheet were shown to be able to prevent sarin gas to pass through at room temperature. However, as shown in Table 1, the number of sarin molecules that passed through the membrane was observed to increase with the temperature. The existence of an attractive resultant potential energy between a sarin molecule and a single porous rGO nanosheet was, then, shown and its energy well, although greater than $k_BT$ at room temperature, was shown not to be greater than the same parameter at temperatures about 900 K. In view of this result, the passage of sarin through two configurations of membranes formed by two adjacent porous rGO nanosheets, as a function of temperature and distance between the nanosheets, was investigated. One configuration consists of a membrane with two rGOs with aligned pores, and the other with non-aligned pores. For these double rGO nanosheet membranes, it was shown that the air filtration from sarin is much more efficient than that of one nanosheet, even at high temperatures, and at different distances between the nanosheets. The membranes also attracted the air molecules, but our results showed that the rGO membranes completely block only sarin ones, thus allowing the passage of air at all temperature values studied here. The possibility to build different membranes from different nanostructures is as wide as the range of conditions to test the filtration of sarin gas. We hope, then, this study might stimulate further research on the filtration of sarin and other chemical warfare gases.

## Acknowledgements

M. A. E. M. acknowledges support from UFSCar and FACENS. A.F.F. is a fellow of the Brazilian Agency CNPq (#311587/2018-6) and acknowledges grant #2020/02044-9 from São Paulo Research Foundation (FAPESP). This research used the computing resources and assistance of the John David Rogers Computing Center (CCJDR) in the Institute of Physics "Gleb Wataghin", University of Campinas.

Air filtration from sarin/air mixture by porous graphene-oxide membranes: a molecular dynamics study

Marco A. E. Maria [a,b,c], Alexandre F. Fonseca [c,*]

[a] Federal University of São Carlos – Sorocaba.

[b] Facens University Center – Sorocaba.

[c] Applied Physics Department, Institute of Physics "Gleb Wataghin", State University of Campinas, Campinas, SP, 13083-970, Brazil.

**Section A – Most significant g(r) curves for epoxide group for 300 K and 17% of sarin concentration.**

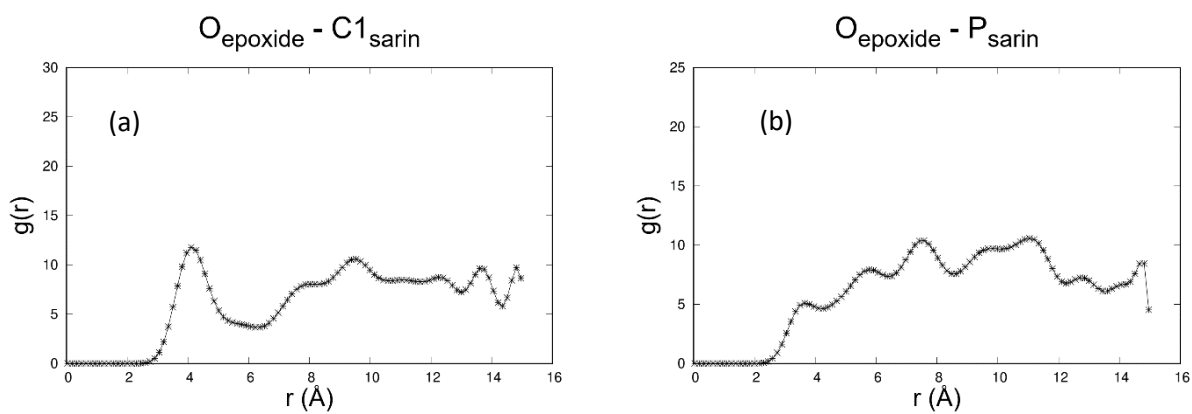

**Figure S1**. Radial distribution function, g(r), between oxygen atom (O) from epoxide group (in the nanosheet) and the atoms C1s (a) and Ps (b), both from sarin molecule.

## Section B – Most significant g(r) curves for hydroxyl group for 300 K and 17% of sarin concentration.

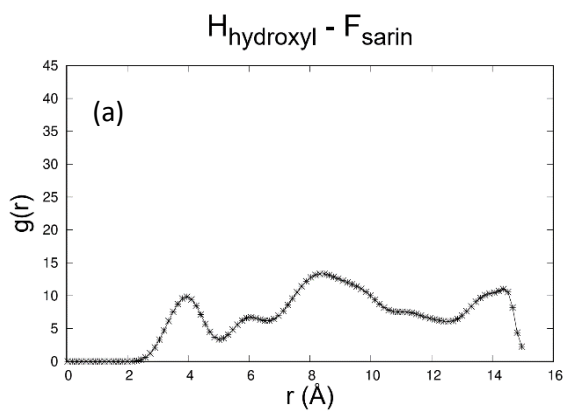
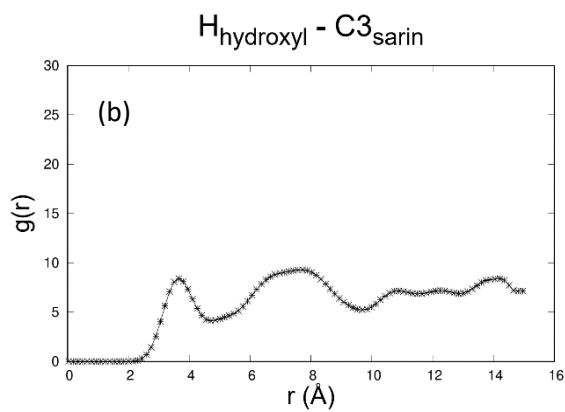
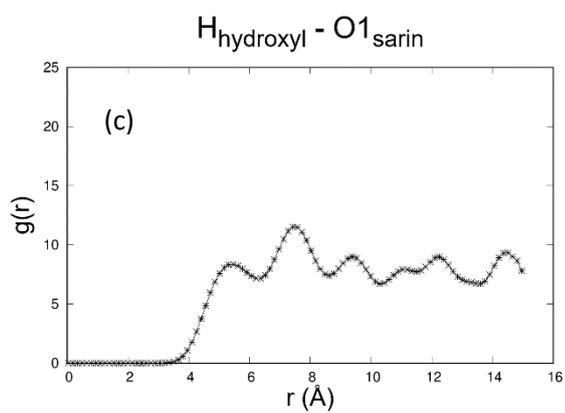
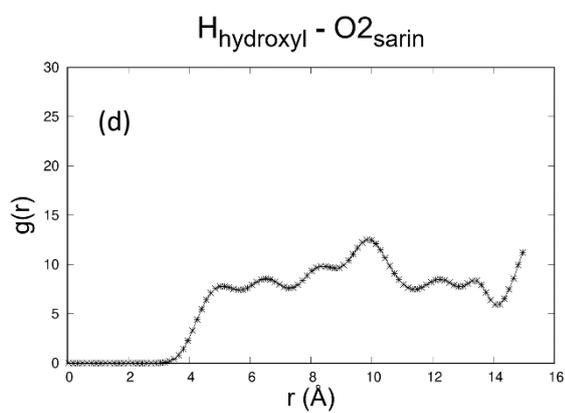
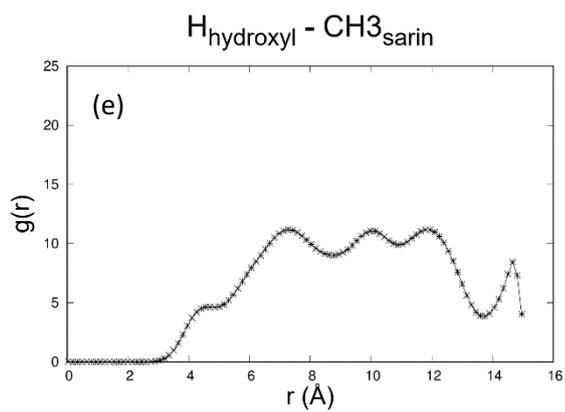
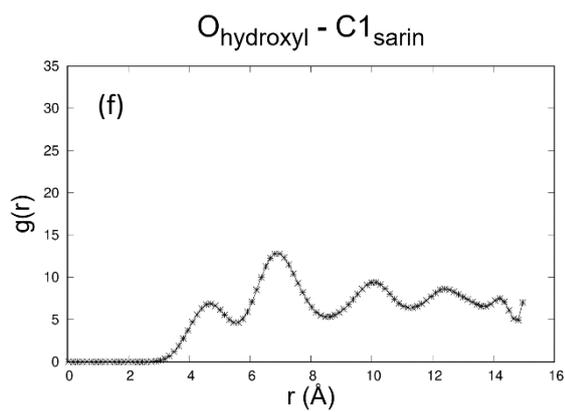

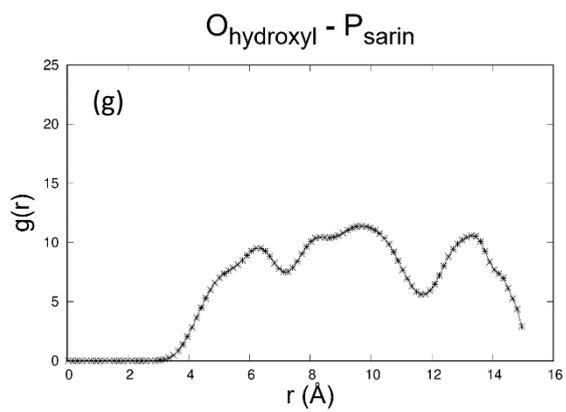

**Figure S2.** Radial distribution function, g(r), between oxygen atom (O) from hydroxyl group (in the nanosheet) and the atoms Fs (a), C3s (b), O1s (c), O2s (d), CH3s (e), C1s (f), and Ps (g), all from sarin molecule.

## Section C – Most significant g(r) curves for carboxyl group for 300 K and 17% of sarin concentration.

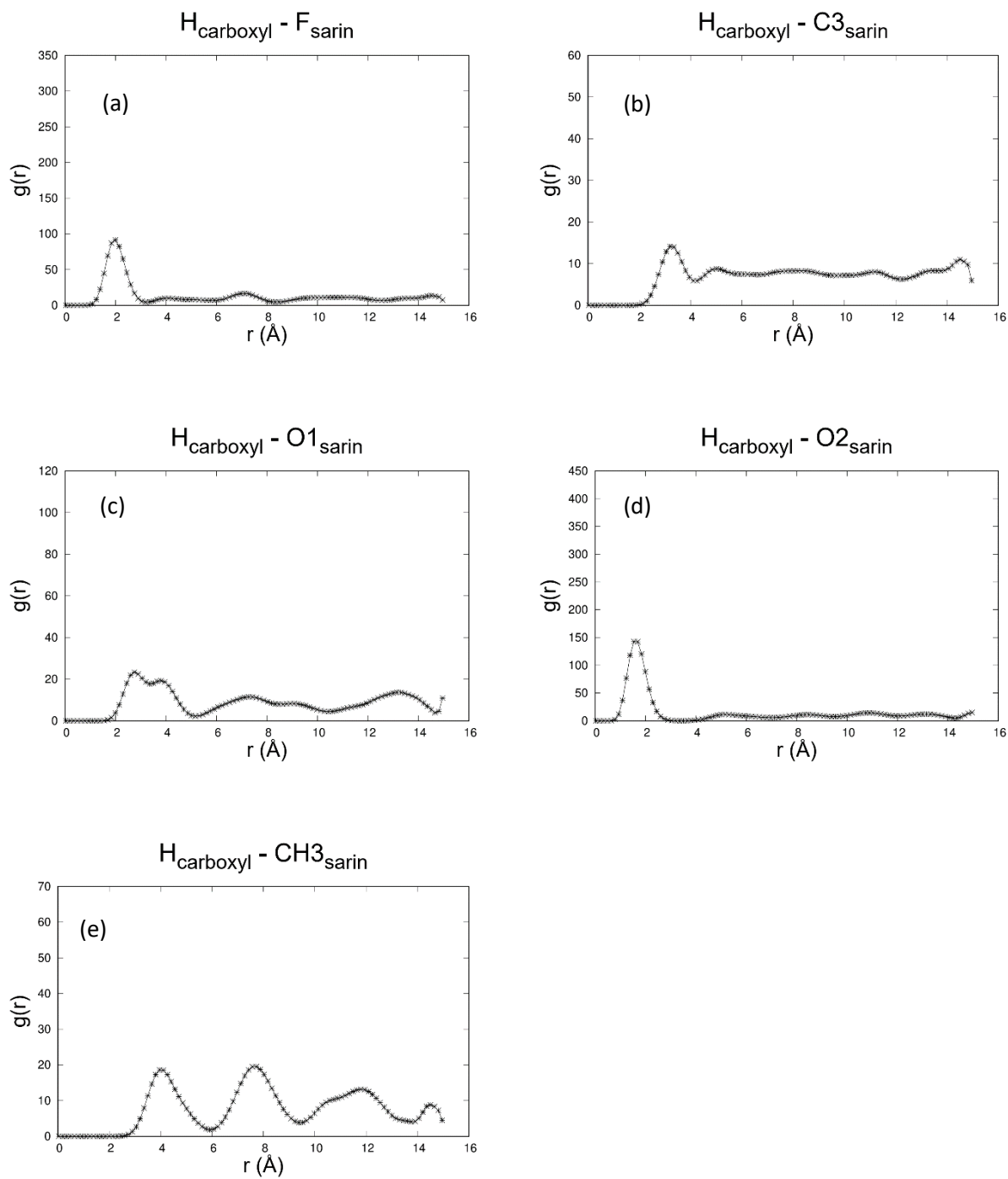

**Figure S3**. Radial distribution function, g(r), between hydrogen atom (H) from carboxyl group (in the nanosheet) and the atoms Fs (a), C3s (b), O1s (c), O2s (d), CH3s (e), all from sarin molecule.

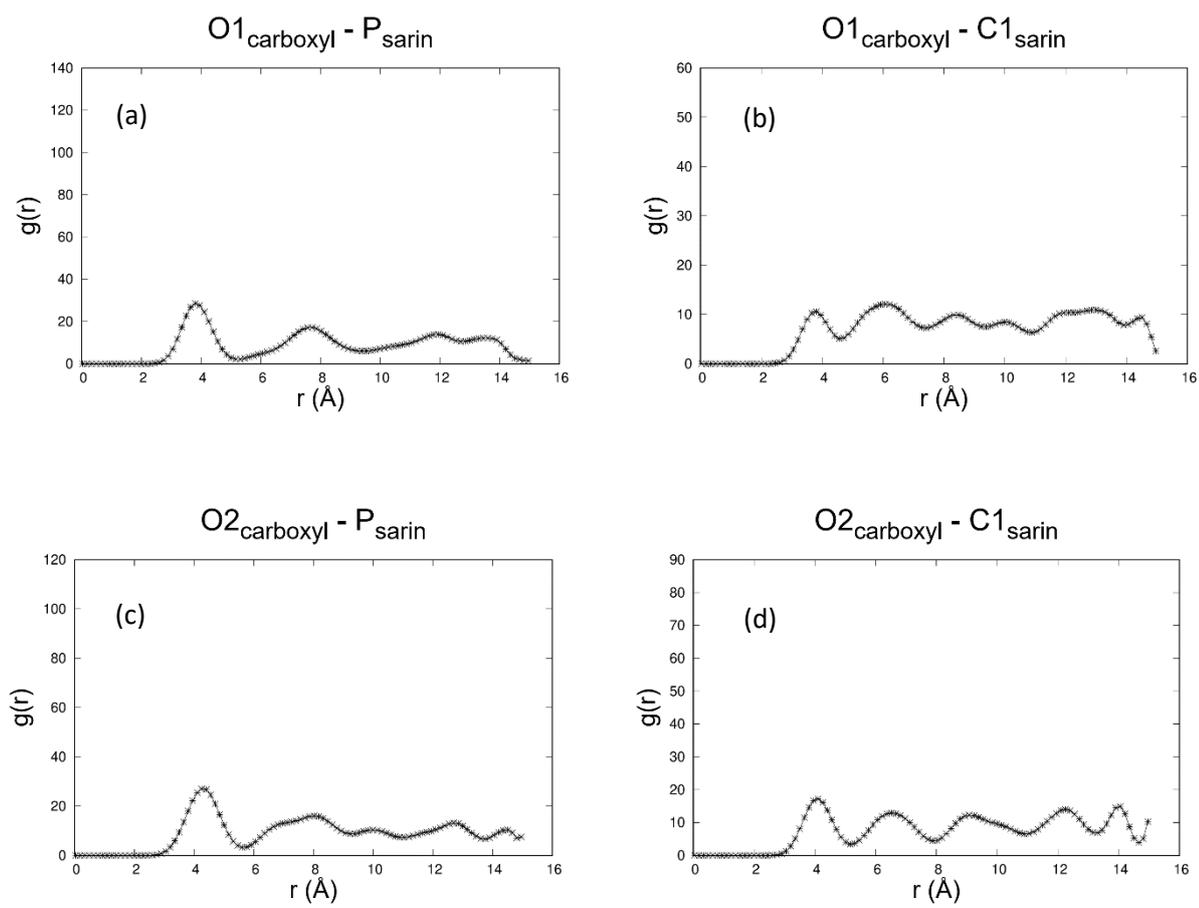

**Figure S4**. Radial distribution function, g(r), between oxygen atom O1 from carboxyl group (in the nanosheet) and the atoms Ps (a), and C1s (b). Radial distribution function between oxygen atom O2 from carboxyl group and the atoms Ps (c), and C1s (d).

# Section D – g(r) curves for nanosheet hydrogens (from carboxyl and hydroxyl) and the O2s, for 17% of sarin concentration.

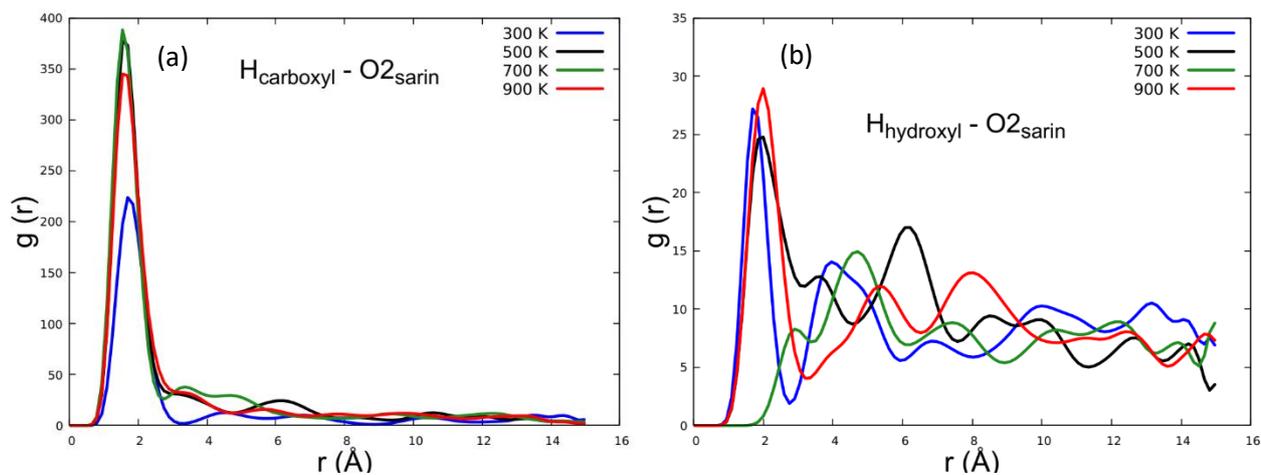

**Figure S5**. Radial distribution function, g(r), curves for hydrogens of carboxyl (a) and hydroxyl (b), both from rGO nanosheet, and the O2s oxygen from sarin, for 17% of sarin concentration in various temperatures.